# Segmenting DNA sequence into 'words'

Wang Liang, Zhao KaiYong

Tencent, SOSO, Beijing, P.R. China

Hong Kong Baptist University, Department of Computer Science, HK, P.R. China

*To whom correspondence should be addressed. E-mail:wangliang.f@gmail.com

**[Abstract]** This paper presents a novel method to segment/decode DNA sequences based on statistical language model. Firstly, we find the length of most DNA "words" is 12 to 15 bps by analyzing the genomes of 12 model species. Then we apply the unsupervised approach to build the DNA vocabulary and design DNA sequence segmentation method. We also find different genomes is likely to use the similar 'languages'.

## 1 Introduction

The letters like A, T, C, G or protein A, R, N is still the basic units to analyze the DNA sequence. Corresponding to English, there are 26 letters. If we only know English letters, we can't give deeply analyzing for English sequences. For example, "iloveapple" only contain few information. We can't connect it to significant terms. It's also difficult to process letter sequence by computer. We need word sequences "I/ love/ apple". Most current information processing systems, such as information retrieval, automatic proofreading, text classification, syntactic parser are all designed in 'words' level, but not letter level.

The English sequence is naturally segmented by space. But for some languages like Chinese, there is no space between letters. The fact that there are no delimiters in sequence posed well know problem of word segmentation. The Chinese sequence is just like "iloveapple", we need segment it into "I/ love/ apple". Segmentation is key step for most Chinese Information Processing (CIP) systems.

DNA sequence is very similar to Chinese. These is also no any space or punctuation. So if we could segment DNA sequence into DNA "words" sequence, we could apply many mature information processing technologies to study DNA sequences. The DNA "words" sequence may also give new hints to discovery the function of DNA.

This paper discuss this problem, how to divide DNA sequence into DNA 'words' sequence. We refer to the Chinese segmentation research and design the DNA segmentation method. Normally, there are two steps in these research. First, we should get a word list or vocabulary. Second, we need design a sequence segmentation method based on vocabulary.

Because we do not have enough linguistic knowledge about DNA sequence, we apply unsupervised methods to build the DNA vocabulary. These methods only need large raw corpus (2,3,4). Fortunately, we have massive amounts of DNA information. This vocabulary building methods will be discussed in paragraph 2. The following paragraph designs the DNA segmentation method. For example, segmentation of "TGGGCGTGCGCTTGAAAAGAGCCTAAG" could be "TCGG/ GC…", "TCGGGC/ GT", etc. We will decide a right segmentation form. The benchmark of segmenting method is also proposed. We give some application of this method in last summary part.

## 2 DNA Vocabulary

**2.1 Experiment data**

Unsupervised methods need large raw DNA sequence. Here we use 12 full genomes of model species as experimental data (Aspergillus, Schizosaccharomyces, Acyrthosiphon, Zebrafish, Strongylocentrotus, Arabidopsis, Caenorhabditis, Fruit Fly, Human, Mouse, Oryza, Xenopus).

**2.1 DNA word length**

To apply the unsupervised method, we should know the maximal length of "words" first. Three "word length" in tripe decoding of gene sequence could only represent 64 kinds of meaning. But the functions of non-gene sequence may be much more complex. "Three word lengths" may be not enough to reveal their linguistic feature. So we guess the DNA words word length is more than 3. Here we apply two statistical methods to evaluate the length of DNA words.

First, we could use "zipf's laws" to evaluate the length of words. The "zipf's laws" states, in a long enough document, about 50% words only occur once. These words are called 'Hapax legomenon'.

Although the DNA sequence is not segmented into words, we could construct words by intersecting segmenting the sequences and calculate the percentage of 'Hapax legomenon'. For instance, the sequence "AAACG", assume the word length is 2. Its intersecting segmentation is AA, AA, AC, CG. There are 3 different words, AA,AC,CG. AA appears twice. AC and CG appears once, which are 'Hapax legomenon'. So there are 2 'Hapax legomenon'. Its percentage is 2/3, about 66%. If for a length, its percentage of "Hapax legomenon" is 50%, we use this length as word length.

The relation of word length 'n' and the percentage of 'Hapax legomenon' in 12 genomes are shown in Fig.1.

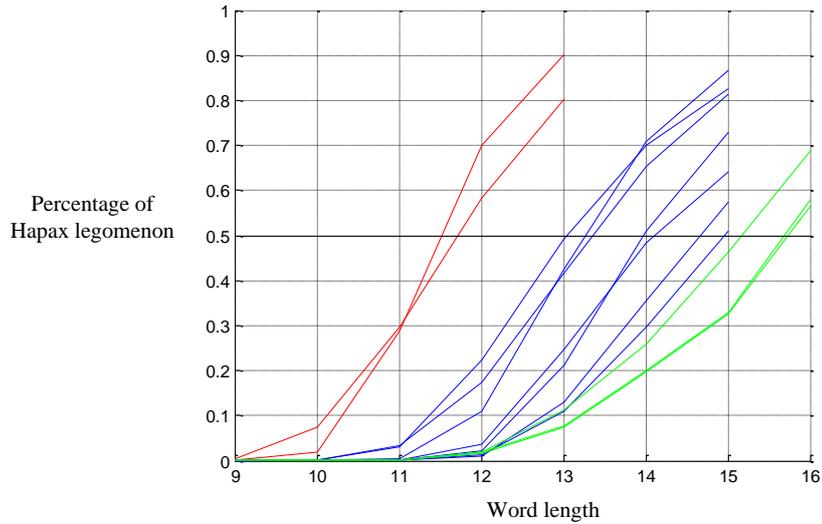

Fig.1 relation of word length and percentage of hapax legomenon (x axis is the word length, y axis is the percentage of hapax legomenon). The red lines correspond to Aspergillus and Schizosaccharomyces. The green lines are Human, Mouse, Xenopus. The other blue lines correspond to other genomes.

In Fig.1, we find 50% line of 'Hapax legomenon' corresponding to word length 12 to 15 of most genomes, which shows the length of most DNA words is no more than 15.

Then we could use the n-grams language model to evaluate the length of words (5). Such model calculates the probability of sequence based on its intersecting segmentation. The n in 'n-grams' means we use word length n to intersecting segment the sequences and build language model.

In n-grams model, we could evaluate the word length n based on this law: the sequence probability will rise with the increase of assumed word length n, till n reach the maximal words length. Normally, we use language perplexity to express the probability of large corpus. This indicator has simple reciprocal relation with probability. So the lowest point of language perplexity will correspond to the maximal words length.

Here we use the 12 genomes to build n-grams models and calculate the language perplexities respectively. The relation of 'n' of "n-gram" and the perplexity of each genome is shown in Fig.2:

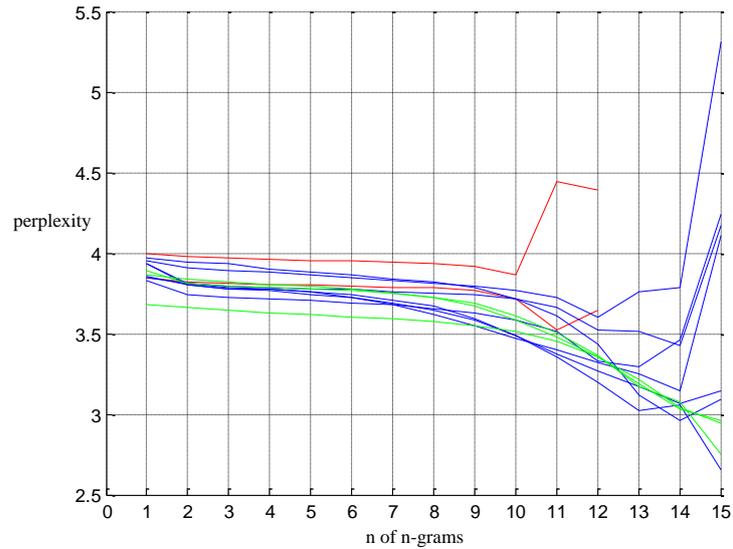

Fig 2. The relation of n-gram 'n' and language perplexity of DNA of model species. The red lines correspond to Aspergillus and Schizosaccharomyces. The green lines are Acyrthosiphon, Zebrafish, Strongylocentrotus. The other blue lines are Arabidopsis, Caenorhabditis, Fruit Fly, Human, Mouse, Oryza, Xenopus.

The Fig.2 shows the language perplexities reduce with the increase of word length n, till n<14. When n > 14 the perplexity of most genomes will increase, which means the language model will not believable for data sparse problem. So we could decide that the upper bound of n of n-gram model for genomes is about 12-15, which shows the fifteen letters almost has no relation with the previous 14 letters in a sequence. It also means the lengths of most words should be no more than 15.

Two methods all show the maximal length of DNA is about 12 to 15. In our experiment, we use the 12 as the maximal length of DNA "words". We should also note that the "word" is a relative concept. For example, the composite word like "big apple" could also be regarded as word. So if we have more corpus, we could set longer length for DNA words.

**2.3 DNA vocabulary**

For 12 word length, we will get $4^1+4^2+….+4^{12}$ = 22,369,620 words. We need evaluate the probabilities of all possible DNA word and then filter the word list to build final vocabulary.

Here we give three methods to get the word probability.

First, we could use the simple frequency method:

Probability of word: P(word) = C(w)/C(N). C(word) is number of word 'w' appear in corpus, C(N) is all word occurrence numbers.

For example: "who is who". C(N)=3, C(who)=2, C(is)=1. So P(who)=2/3, P(is)=1/3. For 2-gram words, C(who is)=1, C(is who)=1, C(N)=2. So P(who is)=1/2, P(is who)=1/2.

Second, we could get the probability of each word based on n-grams language model.

For example, the probability of sequence of "ABC":

P(ABC)=P(A)P(B|A)P(C|AB)

In n-grams language models, we have calculated P(A), P(B|A), P(C|AB) etc. We could directly get all DNA words and its probability. Because language model could apply more smooth method, this probability is more reliable.

Lastly, we could also use EM method to get word probability. For this method, we only give an initial probability for all possible words and then iteratively calculate the probability till convergence (2).

Then we could filter this word list to build final vocabulary according to following rules:

First, The DNA word should have the high frequency. This is a basic ruler to filter the words. For example, in 4-gram counting, the frequency of "love" will be much higher than "ovco", so "love" will be added into vocabulary. "ovco" will be disregarded. Here we select a frequency threshold for 9-12 bps length words. All possible word of 1-9 bps are also added into vocabulary.

Secondly, the connection of letters in word should be strong enough. For example, the frequency of "hisapple" is very high, but the "his" and "apple" is also high frequency word. We could use probability methods to filter these words. The probability P(his)*P(apple) is much more than P(hisapple). So the happening of "hisapple" is only a random collocation, it could not be regard as a word. This rule filters the combination of high frequency words.

Thirdly, the "word" should have clear boundary. For example, 'ur' in 'our' will have a high frequency, but most of its left letter are 'o'. It has no clear left boundary, so it's not a word. Most of such "words" are substring of significant terms. We could use the boundary-verification to eliminate these invalid candidates [10].

After delete these words, we get a vocabulary containing 564,145 words. We use this word set as our DNA vocabulary.

**3 DNA sequence segmentation**

After having a vocabulary with probability, we could use the maximal probability segmentation method to segment the DNA sequence into "DNA words" sequence. It's also mature approaches in natural language processing research. For example, a sequence 'AGC' could be divided into, 'A /G /C', 'AG /C', 'A /GC','AGC'. This method selects a segmentation form having the maximal probability as the segmentation of this sequence.

In segmentation researches, we normally use the precision to measure the effect of a segmentation method. It's the ratio of number of rightly segmenting words to that of all words in the sequence. Since we didn't know the DNA words beforehand, we design a stability indicator to evaluate the effect of DNA sequence segmentation.

For a sequence of "CCCTAAACC", assume its segmentation is "CCC/ TAAA/ C/ C". Then

we delete the first letter of original sequence, the new sub sequence is "CCTAAACC". If its segmentation is "CC/ TAAA/ C/ C", it has one different "word" compared to previous sequence. But if the segmentation is "CCT/ AAA/ CC", it will become a completely different sequence. So a good segmentation method should ensure the sub sequence is segmented into the same form with the original sequence. To run stability test, we only need delete some letters from original sequence and then segment it and calculate the percentage of same segmenting words between this sequence and original sequence.

We randomly select a group of 100 bps length DNA sequences from each genome and run the segmentation stability test by corresponding segmentation models. The average stability ranges from 0.96 to 0.99.

Then we randomly selected 100M data from 12 genomes respectively and create a new vocabulary. The segmentation stability of segmenting method based on this new model ranges from 0.90 to 0.95. This segmentation stability is only 5% lower than previous test, which shows different genomes may share the same vocabulary. So we use this vocabulary built by mixed data as our vocabulary. A single DNA vocabulary for all species will bring many advantages for DNA sequences analyzing.

Here we also discuss an interesting question, all genomes use the same language? We build two dictionaries by rice genome and human genome respectively. Then we use these two dictionaries to segment same sequence. If two dictionaries segment it into same segmented form, they may use the same language.

Just like segment stability metric: first, use two dictionary to segment one sequence. Get two segmented sequences. Then calculate the percentage of same segmenting words between two segmented sequences.

1) Build vocabulary by different chromosomes of human, segment same sequence. Such 'stability ' : about 85%.

2) Build vocabulary by different genomes, segment same sequence. Such 'stability': about 35%--50%.

For data sparse problem, some words only appear several times, its probability is not reliable. So the cross 'stability' above is low. There are mainly tow methods to deal with data sparse problem. First, we could use more sequences/corpus. But single genome data is limited and not enough to evaluate all word prob. Second, we could apply more smooth methods. For example, when we reduce the word length or filter more words, such stability increases.

This result shows that different genomes is likely to use same language.

## 4 Summary and some applications

The DNA 'words' description build a bridge between natural language processing and DNA research. Almost all current text information processing technology could be directly applied in DNA analyzing. Here we use the dictionary built by mixed genomes data.

First, we could find the "hot topic" in genomes. Here we use LDA method to get such topics. Some results are shown as follows:

| Human Chromosome 1 | Topic 0th: TTTTTT, TTTTTTT, GAGAAG, CAACAA, ATATAT<br>Topic 1th: ATATAT, AACAAAA, AATATTT, AACAAA, GAAGGA<br>Topic 2th: AACAAA, GGAGGG, AAGAAA, CTTCCT, TTTTGTT |
|---|---|
| Human Chromosome 2 | Topic 0th: TTTTCT, TTGTTTT, TCTCTC, TTTCTTT, AAGAAA<br>Topic 1th: TTTGTT, TTTTTTT, TTATTTT, TGCCAC, AAACAA<br>Topic 2th: TTTATTT, CTCTCT, ATATAT, ATATTT, CCAGCAG |
| Mouse Chromosome 1 | Topic 0th: CCTCCC, AGGCAG, GAGAGA, CAGGCA, CTGAGGTG<br>Topic 1th: TCTCTCT, AAATAAA, ATAATA, TTTTTCT, AACAAA<br>Topic 2th: AAATAA, AAATAAA, CACACA, GAAGAG, AAGAAAA |

Fig3. Hot topics in some chromosomes

The for alignment method. Current method mainly compare two sequences letter by letter. After segmenting, we could compare them word by word, which will be faster. We could also use the inverted index structure to build a DNA search engine like Google.

Moreover, "automatic proofreading" functions could also be applied in DNA analyzing to check the mutant gene or mistakes in DNA sequencing (6).

The DNA words and related segmentation method give a new description for DNA sequence. If we could find some DNA "words" correspond to biological meaning, it will be the really interesting result.

## Methods and materials

We use the SRILM to build the language model of DNA with Good-turning as discount method (7). All genomes data are downloaded from NCBI (8). The source code of segmentation method of this paper could be found in (9).

**N-gram language model and word length evaluation**

N-grams are sequences of 'n' words in a running text. N-gram frequencies or more sophisticated statistical models of n-gram are widely used for text processing applications such as information retrieval, language identification, etc. In a biological context, n-gram can be sequences of amino acids or nucleotides. For instance, the sequence "AAACG", its unigram are A,A,A,C,G. The 2-grams are AA, AA, AC, CG. Similarly, 3-grams are AAA, AAG, ACG.

N-grams language model uses the basic statistical properties of n-gram. An *n*-grams model predicts $x_i$ based on $x_{i-(n-1)}, \ldots, x_{i-1}$. In Probability terms, this is $P(x_i | x_{i-(n-1)}, \ldots, x_{i-1})$. When used for language modeling, independence assumptions are made so that each word depends only on the last n-1 words. For example, for 3-grams:

$$P(ATCG) \approx P(A)P(T|A)P(C|AT)P(G|TC)$$

The basic statistical feature for an n-grams model is language perplexity or entropy, which describe how well the language model predicts a new text composed of unseen sentences.

The entropy is the average uncertainty of a single random variable:

$$H(X) = -\sum_{x \in X} p(x) \log_2 p(x) \tag{1}$$

For example, the DNA sequence, $x \in \{A, T, C, G\}$, the entropy of one random variable is:

$$-(p(A)\log_2 p(A) + p(T)\log_2 p(T) + p(C)\log_2 p(C) + p(G)\log_2 p(G)) \tag{2}$$

Then for n random variables, corresponding to n length sequence, its entropy:

$$-\sum_{x_i \in X} p(x_1, x_2, \cdots x_n) \times \log_2 p(x_1, x_2, \cdots x_n) \tag{3}$$

For example, the entropy of n=2 DNA length sequence, its entropy:

$$\begin{aligned}-(p(AA)\log_2 p(AA) + p(AT)\log_2 p(AT) + p(AC)\log_2 p(AC) + p(AG)\log_2 p(AG) + \\ p(TA)\log_2 p(TA) + \cdots + p(GG)\log_2 p(GG))\end{aligned} \tag{4}$$

According to Shannon-McMillan-Breiman theorem:

$$H_\infty(X) = \lim_{n \to \infty} \{-\frac{1}{n} \log_2 P(x_1, x_2, \cdots, x_n)\} \tag{5}$$

This value is defined as the language entropy. Its unit is bit. Normally, we use a very long sequence to evaluate this value.

In terms of n-grams analysis, perplexity is a measure of the average branching factor and can be used to measure how well an n-gram predicts the next juncture type in the test set. Perplexity could be calculated by entropy:

$$2^{H(X)} \tag{6}$$

Here we use the method of SRILM to calculate the perplexity. SRILM define the perplexity as:

$$10^{\frac{-\log_{10} P(T)}{Word}}$$, here 'T' is the sequence, 'Word' is the word number in this sequence. This definition has no essential difference to the perplexity definition above.

Because here P(T) is only decided by the word length. The maximal word length evaluation problem could be defined as:

$$Perplexity^* = \arg\min_{WL} (10^{\frac{-\log_{10} P(T|WL)}{Word}}) \tag{7}$$

Here WL is the word length.

**Segmenting method**

Because we have obtained the DNA vocabulary with probability, we could apply the methods from existing research to segment DNA Sequence. One basic method is called 'probability approach to word segmentation'.

We could use an example to show the mission of segmentation. For a sequence "ATAC", assume maximal word length is 3, its segmentation could be "ATA/ C", "AT/ AC", "AT/ A/ C", "A/ TAC", "A/ TA/ C", "A/ T/ AC", " A/ T/ A /C". We select the segmentation candidate which has the maximal probability as the segmentation for the sequence.

In statistical language model, the probability of one form of segmentation is the product of probability of its all words:

$$P^*(ATAC) = \max \begin{cases} P(ATA)P(C) \\ P(AT)P(AC) \\ P(AT)P(A)P(C) \\ P(A)P(TAC) \\ P(A)P(TA)P(C) \\ P(A)P(T)P(AC) \\ P(A)P(T)P(A)P(C) \end{cases}$$

(1)

If the sequence length is m, there will be 2^(m-1) forms of segmentations. To reduce the calculation requirement, dynamic programming methods are applied.

The segmentation problem could be formally defined as:

$S = c_1 c_2 \cdots c_n$ is a sequence of DNA letters.

$W = w_1 w_2 \cdots w_m$ is a sequence of the word segmentation.

What we need is get

$$W^* = \arg\max_W P(W|S) \tag{6}$$

The most probable sequence of segmentation.

According to the Bayes Formula:

$$W^* = \arg\max_W P(W|S) \Rightarrow W^* = \arg\max_W \frac{P(W)P(S|W)}{P(S)} \Rightarrow \arg\max_W P(W)P(S|W) \tag{7}$$

Because the P(S|W) and P(S) are same for all segmentation forms, that leaves us only maxP(W). Based on the words independent assumption, we have:

$$P(W) = \prod_{i=1}^{m} P(w_i) \tag{8}$$

The maximal probability segmentation method obtains a segmentation having maximal P(W). Normally, a word segmentation graph is applied to describe this method. The nodes represent the segmentation positions and the edge is the word with corresponding probability.

For example, a sequence "ATAC", its word segmentation graph is shown in Fig.1.

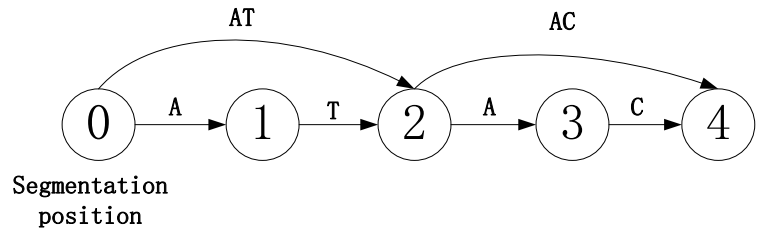

Fig.1. word segmentation graph

In Fig.1, The segmentation positions are 0,1,2,3,4. There are two path from begin node 0 to end node 4.

Path 1: 0---1---2----3----4, its segmentation form "A/ T/ A/ C/", the probability of this segmentation is P(A)*P(T)*P(A)*P(C).

Path 2: 0---2---4, its segmentation form "AT / AC/", its probability is P(AT)*P(AC).

We use segmentation which has the highest probability as the final segmentation form of a sequence. It's a standard optimal route problem. Many dynamic methods could be used to solve this problem.

Here is an example. A sequence in human genome is as follows:

"TGGGCGTGCGCTTGAAAAGAGCCTAAGAAGAGGGGGCGTCTGGAAGGAACCGCAACGCCAAGGGAGGGTG"

Our method will segment it into:

"TGGGCGTG/ C/ G/ CT/ TG/ AAAA/ G/ AGCCT/ AAGAA/ GAGGGGGCGTCTGGA/ AGGAA/ CC/ G/ CA/ A/ C/ GCCA/ AGGGAGGG/ TG/"

## Segmentation stability

For a sequence of "CCCTAAACC", assume two kinds of segmenting methods all divide it into "CCC/ TAAA/ C/ C/".

Then we delete the first letter, it becomes "CCTAAACC":

For the first segmentation method , its segmentation is "CC/ TAAA/ C/ C/".

For the second segmentation method, its segmentation is "CCT/AAAC/C/".

In two segmented sequences, the words having the some begin position and end position could be regarded as the same word. Because we delete the letters from the beginning of sequence, we don't consider the first words in stability calculating.

The first segmentation has 3 same begin/end position pairs with original segmentation, so the stability of the first method is 1. For the second, it has 2 such pairs, but only has one same pair with original segmentation. So its stability is 0.5. This process is illustrated in Fig.3.

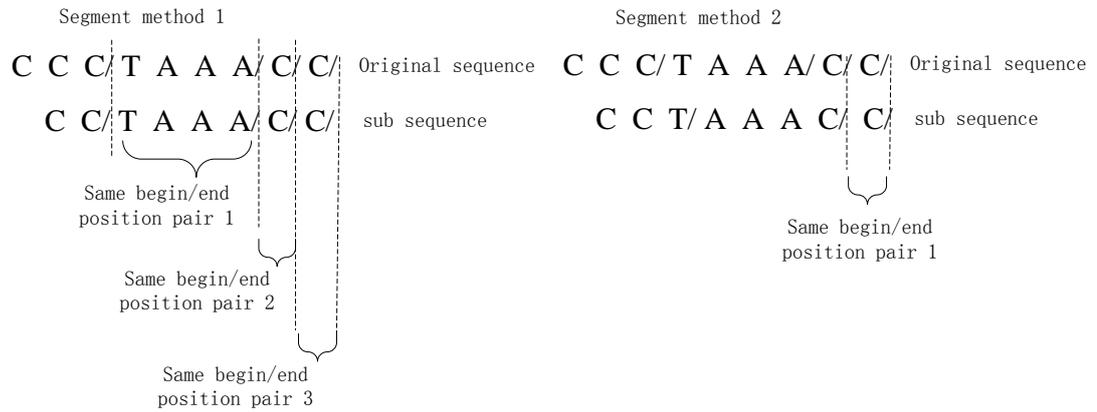

Fig 2. For segment method 1, words number (begin/end position pairs number) in sub sequence is 3, all are same with original sequence , so sability:3/3=1. For method 2, words number is 2, one is same with original sequence, so stability:1/2=0.5.

For vocabulary build by different genomes, the segmentation stability test results are shown in table.1:

Table.1: segmentation stability of different genomes data model for different genomes

| **genomes** | Acyrthosiphon | Arabidopsis | Aspergillus | Caenorhabditis | Zebrafish | Fruit Fly |
|---|---|---|---|---|---|---|
| **stability** | 0.980074 | 0.986467 | 0.973245 | 0.98359 | 0.963535 | 0.983323 |
| **genomes** | Human | Mouse | Oryza | Schizosaccharomyces | Strongylocentrotus | Xenopus |
| **stability** | 0.974546 | 0.965113 | 0.969982 | 0.983754 | 0.970433 | 0.973462 |

The experiments above build different vocabularies for different species. For vocabulary built by mixed genomes data, its segmentation stability is shown in table2.

Table.2: segmentation stability of mixed data model

| **genomes** | Acyrthosiphon | Arabidopsis | Aspergillus | Caenorhabditis | Zebrafish | Fruit Fly |
|---|---|---|---|---|---|---|
| **stability** | 0.942446 | 0.953038 | 0.949611 | 0.933767 | 0.904238 | 0.93521 |
| **genomes** | Human | Mouse | Oryza | Schizosaccharomyces | Strongylocentrotus | Xenopus |
| **stability** | 0.914045 | 0.898843 | 0.909858 | 0.957075 | 0.919044 | 0.92456 |